\documentclass{iaus}
\usepackage{graphicx}

  \checkfont{eurm10}
  \iffontfound
    \IfFileExists{upmath.sty}
      {\typeout{^^JFound AMS Euler Roman fonts on the system,
                   using the 'upmath' package.^^J}%
       \usepackage{upmath}}
      {\typeout{^^JFound AMS Euler Roman fonts on the system, but you
                   dont seem to have the}%
       \typeout{'upmath' package installed. iaus.cls can take advantage
                 of these fonts,^^Jif you use 'upmath' package.^^J}%
      }
  \else
  \fi

  \checkfont{msam10}
  \iffontfound
    \IfFileExists{amssymb.sty}
      {\typeout{^^JFound AMS Symbol fonts on the system, using the
                'amssymb' package.^^J}%
       \usepackage{amssymb}%

      }{}
  \fi

  \IfFileExists{amsbsy.sty}
    {\typeout{^^JFound the 'amsbsy' package on the system, using it.^^J}%
     \usepackage{amsbsy}}
    {\providecommand\boldsymbol[1]{\mbox{\boldmath $##1$}}}

\newsavebox{\astrutbox}
\sbox{\astrutbox}{\rule[-5pt]{0pt}{20pt}}

\title[Outskirts of Galaxy Clusters: intense life in the suburbs]
      {Radio activity in major mergings}

\author[E.Zucca et al.]%
{E.Zucca$^1$, T.Venturi$^2$, S.Giacintucci$^{1,3}$, S.Bardelli$^1$ \and
D.Dallacasa$^3$}

\affiliation{$^1$INAF-Osservatorio Astronomico di Bologna, via Ranzani 1,
40127 Bologna (Italy)\\
e-mail: elena.zucca@bo.astro.it\\[\affilskip]
$^2$ INAF-Istituto di Radioastronomia, via Gobetti 101, 40129 Bologna (Italy)
\\[\affilskip]
$^3$ Dipartimento di Astronomia, Universit\`a di Bologna, via Ranzani 1,
40127 Bologna (Italy)
}
\pubyear{2004}
\volume{195}
\pagerange{1--8}
\date{?? and in revised form ??}
\jname{Outskirts of Galaxy Clusters: intense life in the suburbs}
\editors{A. Diaferio, ed.}
\begin{document}
\maketitle

\begin{abstract}
Cluster mergings are expected to have an influence on the radio emission
of the galaxy population. We present the results of a deep radio survey in the 
A3558 complex in the central region of the Shapley Concentration, in order to 
further explore our hypothesis of a dependence of the radio luminosity 
function on the age of the merging.
\end{abstract}

\section{Introduction}

During the huge energy release of a cluster merging, different mechanisms
are at work on galaxies. On one side, the rapid changes in the tidal 
gravitational field can drive gas from outer to inner parts
of galaxies, triggering star bursts and feeding the central AGN (\cite{bekki}),
while on the other side tidal stripping and ram pressure tend to devoid
galaxies of their hydrogen.

\cite{owen99} found that A2125, a merging cluster, presents a factor 4 more
radiogalaxies than a relaxed one (A2645) at the same distance.
In order to better study this phenomenon, we performed an extensive
radio survey at 20cm with the ATCA telescope on the cluster complexes
(\cite{bardelli98a}, \cite{bardelli98b}, \cite{bardelli00}, \cite{bardelli02})  
located in the Shapley Concentration (\cite{zucca93}). 

The only way to estimate the radio activity of the clusters is to compute
the bivariate radio-optical luminosity function (Figure 1), i.e. the number of
radiogalaxies in the considered structure (as confirmed by their redshift) over 
the total number of optical galaxies. This takes into account the different 
richness of the clusters.
We found that the A3558 complex, which is a merging in an advanced phase, 
presents a significant lack of radiosources (\cite{venturi00}, 
\cite{venturi01}, \cite{venturi02}). 
On the contrary the A3528 complex, which is at an early phase of interaction,
presents a number of radiosources comparable to that of normal clusters.
Also the A3571 complex, a merging at the end of the process, does not have 
pecularities in the number of radio objects.

This result is only apparently in contraddiction with \cite{owen99}, because
one has to take into account the spiral richness of interacting clusters and/or
the merging phase. Our proposed scenario is the following:
at very early stages of merging, the outer parts of clusters (which are spiral 
rich) begin to interact, triggering star formation activity and therefore
enhancing the number of radiosources. After this episode, the radiosource 
excess disappears and the clusters return quiescent (right panel of Figure 1),  
until the ellipticals (with their AGNs) in the inner part of the clusters are
reached by the merging effect, leading to a lowering of the AGN emission: as a 
result, there is a deficit of bright radiosources (left panel of Figure 1).
Finally, the virialization re-establishes the starting situation.

\begin{figure}
\centering
\includegraphics[angle=0,width=0.5\hsize]{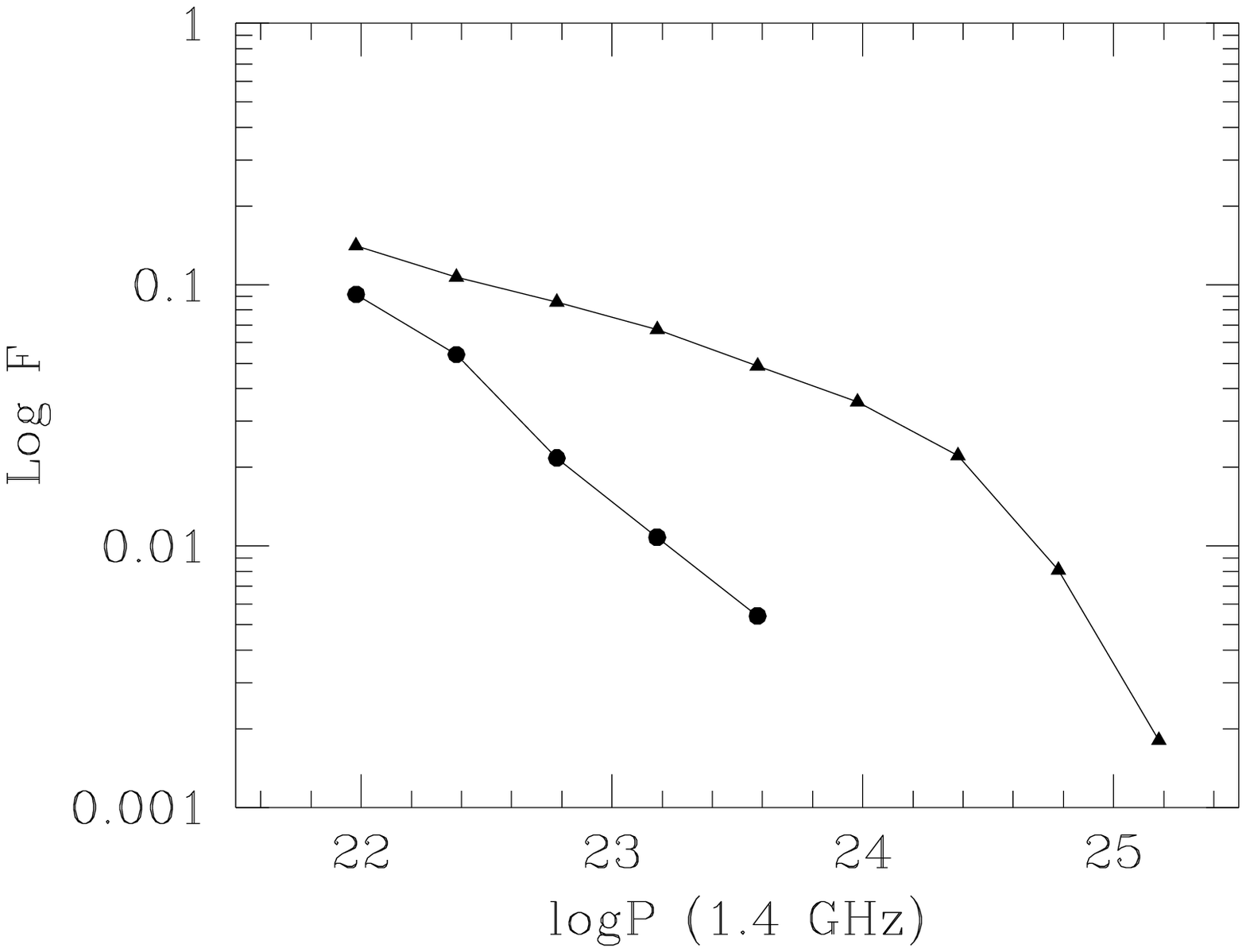}
\includegraphics[angle=0,width=0.4\hsize]{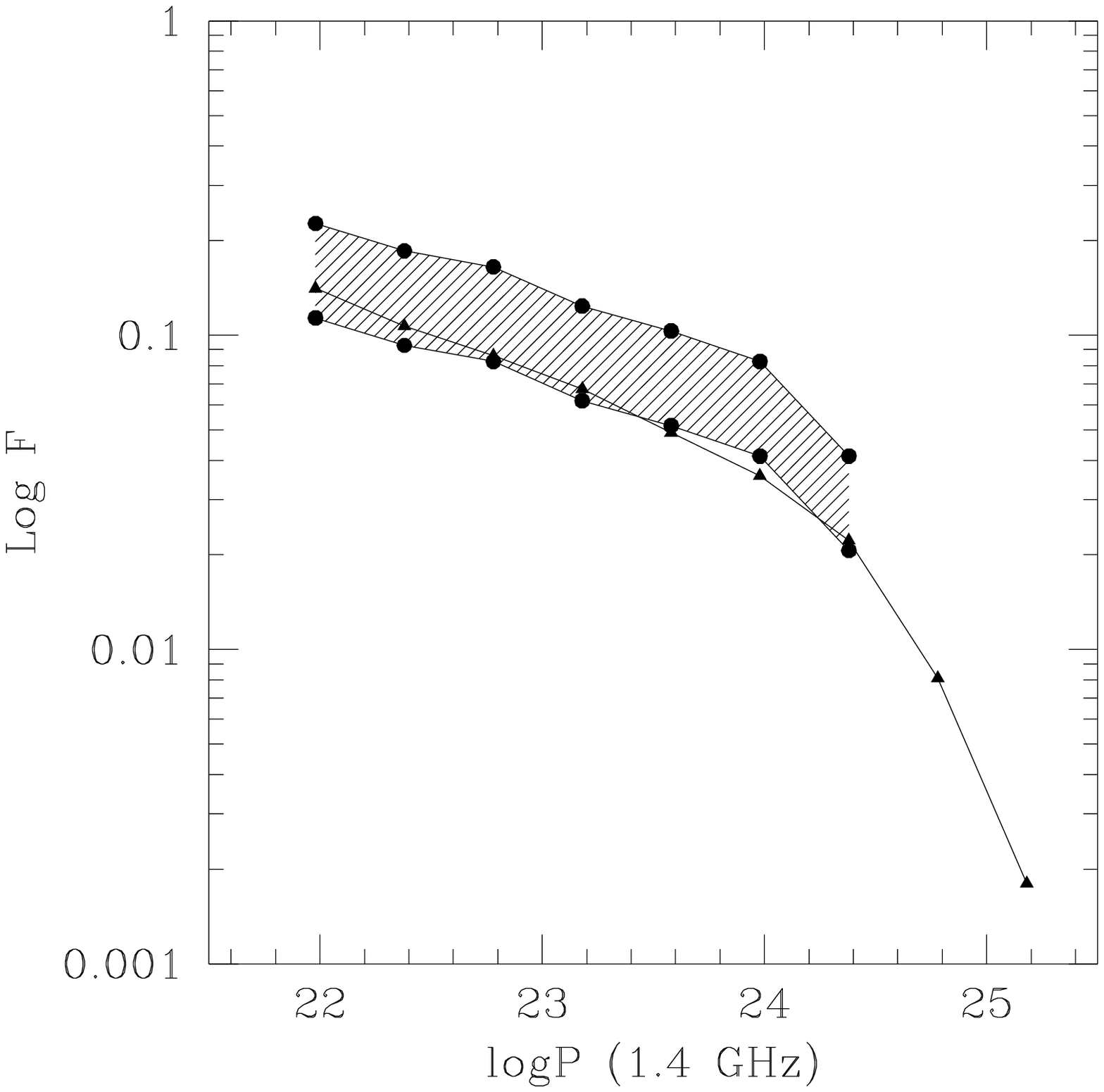}
\caption{Bivariate radio-optical luminosity functions of the A3558 complex
(left) and of the A3528 complex (right). Triangles represent the reference 
luminosity function of normal clusters from \cite{lo96}. 
The first complex is an advanced merger, while the second is a merging at 
early phases. For the A3528 complex, the radio luminosity function lies in the 
shadowed area.}
\end{figure}

\begin{figure}
\centering
\includegraphics[angle=0,width=0.7\hsize]{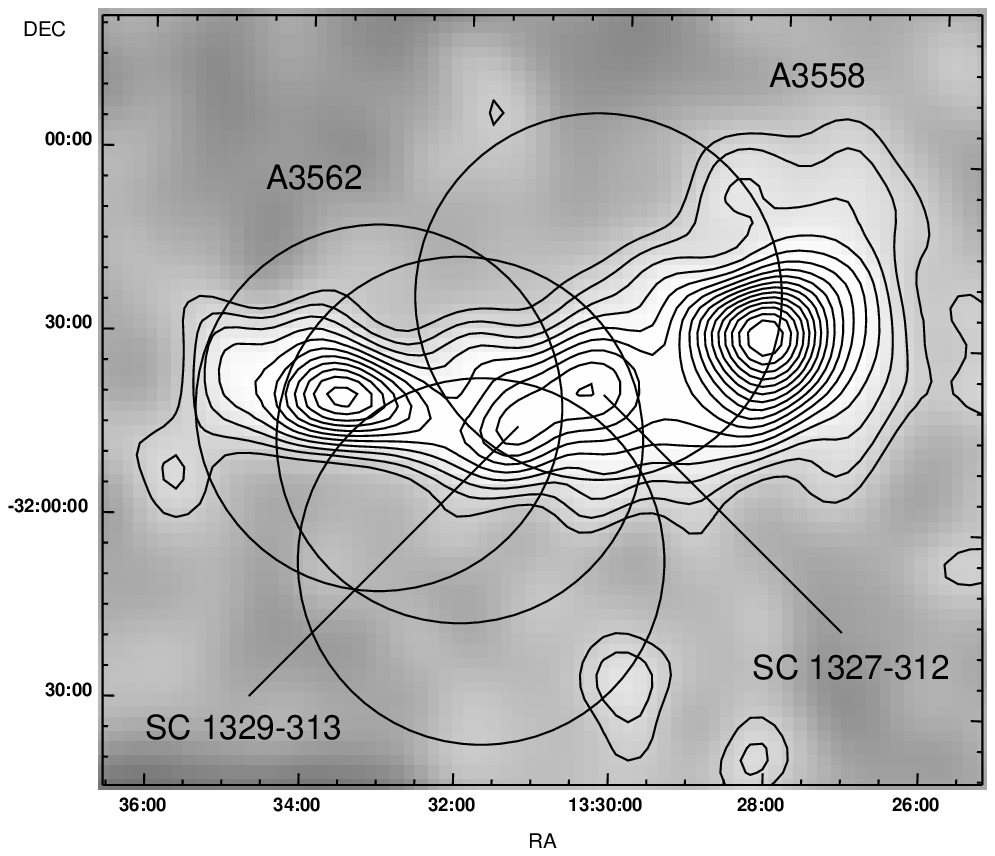}
\centering
\includegraphics[angle=0,width=0.5\hsize]{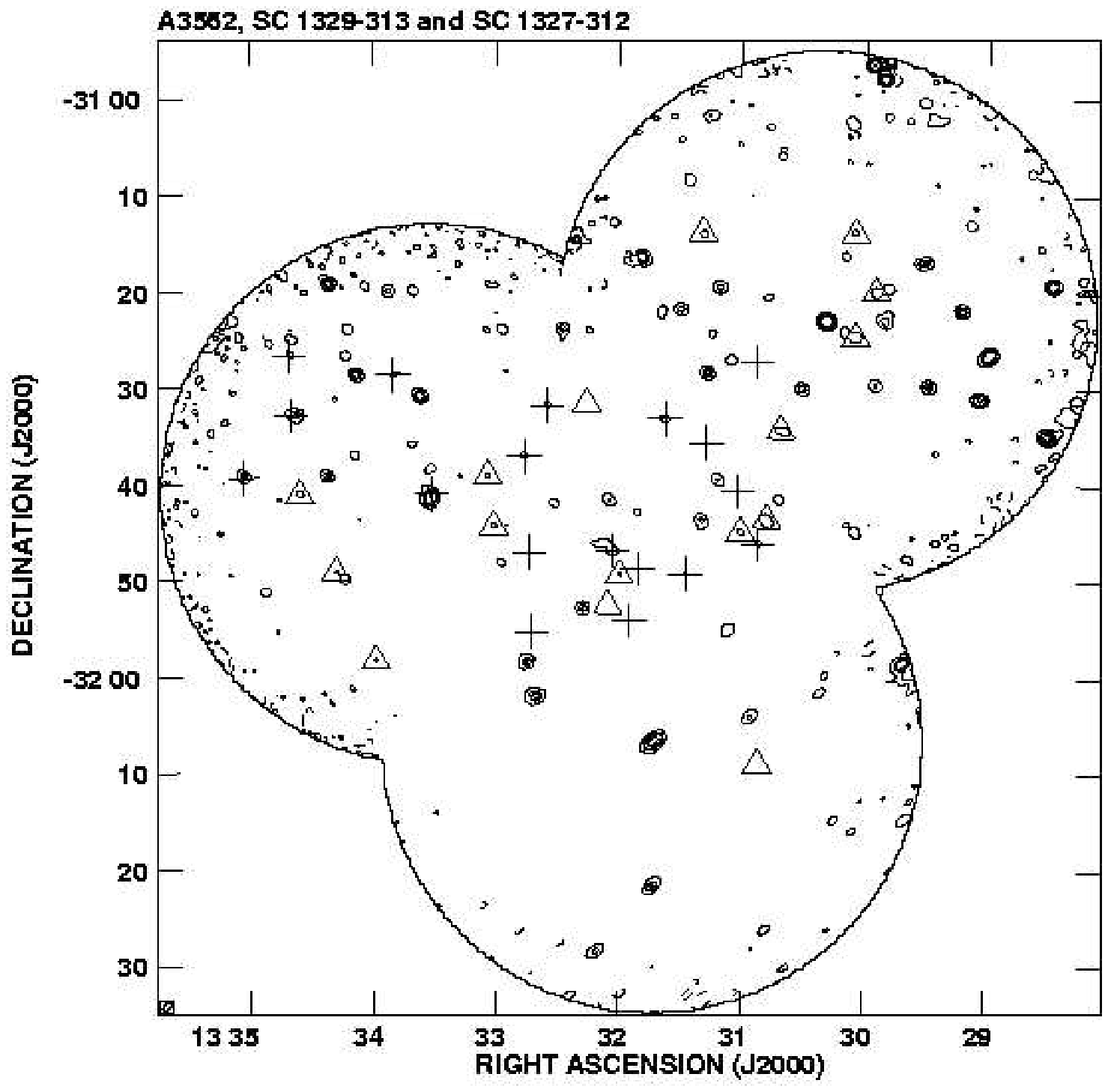}
\caption{{\it Upper panel:} circles represent the field of view of our VLA 
20cm pointings, superimposed on the isodensities of the optical distribution 
of galaxies in the A3558 complex. {\it Lower panel:} location of optical 
counterparts overplotted to the radiocontours. 
Triangles are late type galaxies, crosses are early type galaxies.}
\end{figure}

\section{The deep VLA survey}

In order to better study the lack of radiosources in the A3558 complex,
we performed a dedicated, deep ($\sim 0.25$ mJy b$^{-1}$) survey in the region 
between the two clusters A3558 and A3562 (see Figure 2, \cite{giacintucci04}). 
This region is promising because it is highly substructured and presents 
features indicating an ongoing merging (see the contribution of Bardelli,
this conference). Moreover, we discovered an interesting radio halo 
in the A3562 cluster (see the contribution of Giacintucci, this conference). 
The groups of this region tend to have bluer galaxies, indicating a possible 
ongoing starburst, and their hot gas distributions are highly disturbed
(\cite{bardelli02}).

Given the high statistics, we computed the radio luminosity function for both
early and late type galaxies in this region, finding for elliptical and S0 
galaxies consistency with \cite{lo96}.
Given the lack of bright radiosources that we found analysing the whole
A3558 complex (see previous section), this result suggests that
the deficit is entirely due to the A3558 cluster. 
Moreover, a population of faint ($logP_{1.4 GHz}[W\ Hz^{-1}]<22$) radiogalaxies 
has been found: half of these objects are blue, suggesting that starburst is 
the driving radio emission mechanism.
Finally, we found 14 spiral galaxies, whose ratio between radio and optical
emission is similar to that of galaxies located in rich and dynamically evolved 
clusters.

The next steps of this analysis will be the following:
\\
1) extending the study to other systems in other superclusters: in particular
we are analysing the cluster A2061 in the Corona Borealis supercluster;
\\
2) extending the survey to other wavelengths: in particular we obtained
$610$ MHz observations of the A3558 complex with the GMRT telescope;
\\
3) studying the HI emission of spiral galaxies in order to see whether the 
ongoing merging in the A3558 complex was able to strip gas from galaxies in 
between the dominant clusters.



\end{document}